\documentclass[lettersize,journal]{IEEEtran}
\usepackage{enumitem} 
\usepackage{amsmath,amsfonts}
\usepackage{orcidlink}
\usepackage{algorithmic}
\usepackage{algorithm}
\usepackage{array}
\usepackage[caption=false,font=normalsize,labelfont=sf,textfont=sf]{subfig}
\usepackage{textcomp}
\usepackage{stfloats}
\usepackage{url}
\usepackage{verbatim}
\usepackage{graphicx}
\usepackage{cite}
\usepackage{float}
\usepackage{xcolor}

	\definecolor{darkblue}{RGB}{0,0,150}  

\begin{document}

\title{SleepLiteCNN: Energy-Efficient Sleep Apnea Subtype Classification with 1-Second Resolution Using Single-Lead ECG}

\author{Zahra Mohammadi \orcidlink{0009-0005-0450-8367}, and Siamak Mohammadi \orcidlink{0000-0003-1515-7281},~\IEEEmembership{Senior Member,~IEEE}
 \thanks{Z. Mohammadi is a Ph.D. Candidate in the School of Electrical and Computer Engineering, University of Tehran, Tehran, Iran (e-mail: zahramohammmadi@ut.ac.ir).}

 \thanks{S. Mohammadi is an Associate Professor in the School of Electrical and Computer Engineering, University of Tehran, Tehran, Iran (e-mail: smohamadi@ut.ac.ir).}
}

\markboth{}%
{Shell \MakeLowercase{\textit{et al.}}: A Sample Article Using IEEEtran.cls for IEEE Journals}


\maketitle

\begin{abstract}Apnea is a common sleep disorder characterized by breathing interruptions lasting at least ten seconds and occurring more than five times per hour. Accurate, high-temporal-resolution detection of sleep apnea subtypes—Obstructive, Central, and Mixed—is crucial for effective treatment and management. This paper presents an energy-efficient method for classifying these subtypes using a single-lead electrocardiogram (ECG) with high temporal resolution to address the real-time needs of wearable devices. We evaluate a wide range of classical machine learning algorithms and deep learning architectures on 1-second ECG windows, comparing their accuracy, complexity, and energy consumption. Based on this analysis, we introduce \textit{SleepLiteCNN}, a compact and energy-efficient convolutional neural network specifically designed for wearable platforms. SleepLiteCNN achieves over 95\% accuracy and a 92\% macro-F1 score, while requiring just 1.8~$\mu$J per inference after 8-bit quantization. Field Programmable Gate Array (FPGA) synthesis further demonstrates significant reductions in hardware resource usage, confirming its suitability for continuous, real-time monitoring in energy-constrained environments. These results establish SleepLiteCNN as a practical and effective solution for  wearable device sleep apnea subtype detection. The implementation code is publicly available at:  \href{https://github.com/zahraaayii/SleepLiteCNN-Energy-Efficient-Sleep-Apnea-Subtype-Classification}{github.com/zahraaayii/SleepLiteCNN-Energy-Efficient-Sleep-Apnea-Subtype-Classification}.
\end{abstract}


\section{Introduction}

\IEEEPARstart{A}{pnea} is a common sleep disorder in humans and animals, characterized by interrupted or reduced breathing for at least ten seconds and occurring more than five times per hour \cite{1}. Symptoms include snoring, daytime fatigue, and headaches, while complications may lead to hypertension, memory issues, and stroke \cite{1},\cite{2}. Sleep apnea has three types: obstructive (OSA), central (CSA), and mixed (MSA). OSA occurs when throat muscles relax, blocking the upper airway and reducing breathing efficiency \cite{1}. CSA results from the brain and nervous system failing to send respiratory signals \cite{1}. MSA combines features of both OSA and CSA \cite{1}. Accurate detection of these subtypes is crucial for effective treatment and management strategies \cite{3},\cite{4}.

The standard method for diagnosing sleep apnea relies on polysomnography (PSG), where patients sleep in a clinical setting while multiple physiological signals—electroencephalogram (EEG), ECG, blood oxygen saturation (SpO\textsubscript{2}), and snoring—are recorded and analyzed by sleep experts \cite{5}. Although PSG is highly accurate, it is costly, time-consuming, and uncomfortable for patients, creating a growing demand for alternative diagnostic methods. Since apnea and the associated reduction in breathing significantly impact ECG patterns \cite{6}, machine learning and deep learning applied to wearable ECG-based systems offer a promising alternative.

Despite advancements, most existing studies focus only on binary classification—distinguishing normal breathing from apnea (or just OSA)—and rely on relatively large temporal resolutions (e.g., 30 seconds or 1 minute), which are impractical for real-time or near-real-time feedback in wearable devices. In contrast, we propose classifying all sleep apnea subtypes (normal, OSA, CSA, and MSA) with a 1-second resolution, enabling dynamic therapeutic interventions such as continuous positive airway pressure (CPAP) adjustments to improve patient outcomes. This finer temporal resolution is crucial for prompt interventions and continuous home monitoring in wearable platforms.

Wearable health-monitoring devices face significant constraints on energy consumption, memory size, and computational complexity. Conventional deep neural networks, while highly accurate, often have large parameter counts, leading to increased power consumption, shorter battery life, and limited practicality on resource-constrained wearable hardware. To systematically analyze these limitations, we comprehensively evaluate six classical machine-learning algorithms and eight state-of-the-art deep-learning architectures, examining their performance, complexity, and energy efficiency using identical 1-second windows of single-lead ECG data.

Drawing on insights from this detailed benchmarking, we introduce SleepLiteCNN, a lightweight convolutional neural network(CNN) designed specifically for energy-efficient sleep apnea subtype classification. SleepLiteCNN achieves near state-of-the-art accuracy (95\%) and macro-F1 score (92\%), while substantially reducing energy consumption to 1.8~$\mu J$ per inference after quantization. Furthermore, FPGA hardware synthesis demonstrates its suitability for practical wearable deployment, confirming significant resource savings compared to standard deep neural networks.

Specifically, the main contributions of this paper include:

\begin{enumerate}
	\item \textbf{SleepLiteCNN: Energy-efficient, accurate apnea classification.}  
	We introduce a novel CNN architecture optimized explicitly for wearable applications, delivering strong accuracy while maintaining ultra-low energy consumption.
	
	\item \textbf{Systematic evaluation and benchmarking.}  
	We comprehensively compare classical machine-learning methods and recent deep-learning models using single-lead ECG, thoroughly assessing each model’s accuracy, macro-F1 score, parameter complexity, and measured energy per inference.
	
	\item \textbf{Energy-aware quantization analysis.}  
	We implement 8-bit quantization across all deep-learning architectures, providing detailed, realistic estimates of energy consumption and demonstrating SleepLiteCNN’s significant advantages for wearable scenarios.
	
	\item \textbf{FPGA hardware synthesis validation.}  
	To demonstrate practical feasibility, we synthesize SleepLiteCNN onto FPGA hardware, verifying that it achieves notable resource savings, making it highly suitable for deployment in  resource-constrained wearable environments.
	
	\item \textbf{Real-time, subtype-level apnea classification.}  
	SleepLiteCNN efficiently discriminates between normal breathing, OSA, CSA, MSA at 1-second resolution, enabling real-time feedback and continuous monitoring through wearable technology.
\end{enumerate}

The remainder of this paper is structured as follows: Section II reviews related work on sleep apnea classification and energy-efficient machine learning. SectionIII describes our methodology, including data preprocessing, feature extraction, and model development. SectionIV presents experimental results with a comprehensive analysis. SectionV discusses limitations and future research directions, while Section~VI concludes the paper.

\section{Literature Review}

This section reviews prior research on sleep apnea detection across four key areas: (1) ECG-based classification scope (binary vs. full subtypes), (2) temporal resolution for apnea event detection, (3) traditional machine learning vs. deep learning approaches, and (4) energy-efficient designs with hardware considerations. Finally, we identify existing gaps and provide a detailed comparative analysis.

Numerous studies have explored ECG-based sleep apnea detection, yet most focus solely on binary classification, distinguishing normal breathing from apnea \cite{7,8,9,10,11,12,13,14}. However, identifying specific subtypes is essential for targeted treatment \cite{3,4}. While some recent studies have attempted multiclass classification, they often rely on non-ECG signals and fail to address real-time, low-power requirements. For instance, \cite{2} used EEG from the Tianjin dataset to classify normal breathing, OSA, and CSA, whereas \cite{15} utilized multiple EEG datasets for subtype identification but reported only moderate performance (81.39\% accuracy, 56\% F1-score) on the St. Vincent's University Hospital / University College Dublin Sleep Apnea Database (UCDDB).

Historically, researchers have used 30-second or 1-minute segments to align with conventional PSG scoring. While these larger segments help capture stable signals, they are unsuitable for real-time feedback in wearable applications. Recent studies have shifted toward shorter windows to improve responsiveness. For example, \cite{10} employed a 5-second window, achieving 79.61\% accuracy in hypopnea/apnea classification, whereas \cite{7,8,9} processed signals using 11-second windows for a 1-second resolution, incorporating pruning and binarization to reduce complexity.

Traditional machine learning approaches remain widely used due to their interpretability and lower computational overhead, as demonstrated in \cite{12}, which compared fourteen machine learning and nineteen deep learning models on the Apnea-ECG Database (AED) \cite{6}. The best-performing hybrid model, ZFNet-BiLSTM, achieved 88.13\% accuracy and an 84.04\% F1-score for binary classification. Other machine learning studies \cite{2,13,15} have reported competitive results through feature engineering; however, deep learning has gained momentum for its ability to automatically extract robust features from raw inputs \cite{9,8,16}.

Energy efficiency in sleep apnea detection has been explored through various approaches, including reducing computational overhead and optimizing power consumption. For instance, \cite{24} and \cite{25} implemented FPGA-based solutions using non-ECG signals (EEG and SpO$_2$), but their focus remained limited to binary classification. Meanwhile, \cite{7,8,9} reported energy consumption for a pruned 1D-CNN implemented on a 28~nm node. However, none of these approaches address full-subtype classification using ECG signals—a gap that our work fills by providing comprehensive ECG-based classification with enhanced energy optimization.

A closer examination of the literature shows that ECG-based methods predominantly focus on binary classification, while attempts at multiclass detection often rely on longer windows, inconsistent results, or non-ECG signals. In addition, most short-window approaches fail to address energy constraints or systematically assess real-world resource usage. As our detailed comparative analysis reveals, these gaps underscore the need for a solution that combines finer temporal granularity with robust subtype detection and hardware-aware optimization. Our work meets this need by offering full-subtype classification (OSA, CSA, MSA, normal) at a 1-second resolution, integrating energy-consumption measurements and quantization for efficiency, and validating the design on FPGA—an approach well-suited for wearable, real-time sleep apnea monitoring.

\section{Methodology}

\begin{figure*}[!t]
	\centering
	\includegraphics[width=1\linewidth]{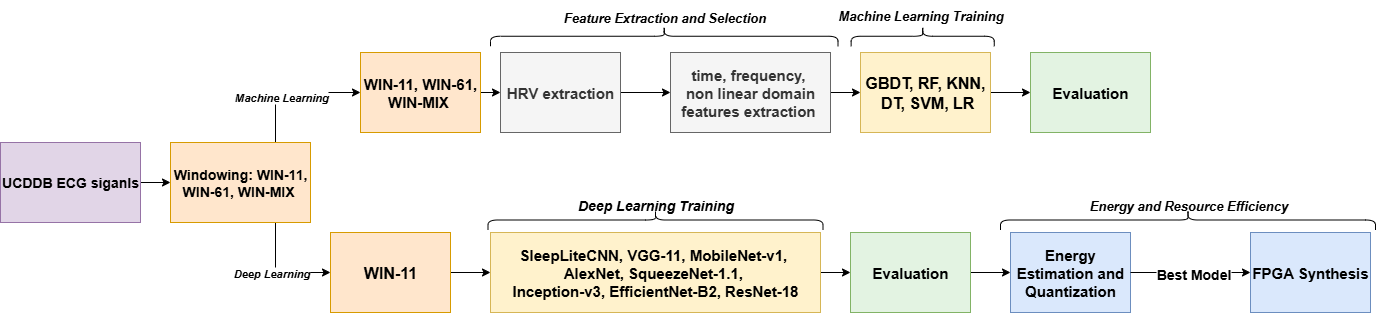}
	\caption{End-to-end pipeline for energy-aware 1-second sleep-apnea-subtype classification. Raw UCDDB ECG signals are segmented with three windowing schemes (WIN-11, WIN-61, WIN-MIX). The upper branch extracts HRV features and trains classical models; the lower branch trains  lightweight deep networks and our proposed SleepLiteCNN. After evaluation, the best deep model undergoes 8-bit quantization, energy estimation, and FPGA synthesis for real-time wearable deployment.}
	\label{fig_1}
\end{figure*}

As illustrated in Fig.~\ref{fig_1}, our methodology focuses on high temporal resolution and energy efficiency for wearable ECG-based sleep apnea subtype classification. We implement a windowing strategy to achieve a 1-second resolution and compare traditional machine learning with deep learning approaches. In the machine learning pipeline, we extract and select handcrafted features from ECG windows before training classical algorithms with cross-validation. For deep learning, we evaluate several lightweight architectures and introduce a SleepLiteCNN optimized for 1D ECG signals, significantly reducing parameters and energy consumption while maintaining robust performance. To further enhance energy efficiency, we apply 8-bit quantization and estimate energy consumption for each deep learning model. Finally, we synthesize the most efficient model on an FPGA to validate its feasibility for continuous operation on battery-powered devices. The following subsections provide a detailed breakdown of each stage, from windowing and feature extraction to model design.

\subsection{Dataset}

This paper utilizes the UCDDB dataset, which contains 25 full overnight PSG recordings (6–8 hours each) from adult subjects suspected of sleep-disordered breathing. These recordings, annotated by sleep experts at a 1-second resolution, include multiple physiological signals, though we focus on a single-lead ECG (modified lead V2) sampled at 128 Hz \cite{17}. Since apnea and hypopnea share similar physiological characteristics, we classify them as a single condition \cite{6}. After applying data windowing, we split the dataset into training (80\%) and testing (20\%) sets. From the training set, we further allocate 10\% as a validation set for use in deep learning model development. The test set is reserved exclusively for final performance evaluation to prevent data leakage. For machine learning models, we perform 5-fold cross-validation within the training subset during Bayesian hyperparameter tuning.

\subsection{Windowing}  
To achieve a 1-second resolution for sleep apnea classification while meeting the ten-second duration constraint for apnea events, we applied a windowing technique to segment ECG signals. We developed three windowing approaches: WIN-11, WIN-61, and WIN-MIX. Our analysis of the UCDDB dataset confirmed no signal loss in the ECG recordings, allowing us to utilize all available signals without requiring gap-filling. Additionally, we opted not to denoise or normalize  the ECG signals, enabling our models to learn noise-resilient features and improving their generalization to real-world wearable applications.
\begin{itemize}
	\item{WIN-11: Uses an 11-second window, with the first second serving as the label and 1-second resolution, the remaining ten seconds ensuring compliance with the apnea event duration constraint. This approach is used for deep learning models..}
	\item{WIN-61: Employs a 61-second window, where the first second is the label, and 1-second resolution, followed by 60 seconds of ECG data. This longer duration allows for the extraction of a comprehensive set of features, including time domain, frequency domain, and nonlinear domain features. The extended window is particularly crucial for accurately computing frequency domain and nonlinear domain features, which require at least 60 seconds of ECG signal \cite{19}.}
	\item{WIN-MIX: A hybrid approach that integrates both WIN-11 and WIN-61, using an 11-second window for time-domain features and a 61-second window for frequency-domain and nonlinear-domain feature extraction. This method balances detailed feature extraction with minimal signal overlap.
	}
\end{itemize}
For our machine learning models, we used all three windowing techniques, as they allow for the extraction of different numbers and types of features from varying signal durations. This multi-windowing approach provides a richer feature set for the machine learning algorithms to work with, potentially enhancing their classification performance.
\subsection{Undersampling}
The UCDDB dataset exhibits a significant class imbalance, particularly in sleep apnea subtype classification. As shown in Table~\ref{tab:Undersampling}, only 10\% of all ECG windows correspond to apnea subtypes, which can introduce bias into classification models. Dataset balancing techniques include undersampling, oversampling, and augmentation. In this study, we applied undersampling to achieve a more balanced dataset. Through iterative testing and performance evaluation, we determined the optimal number of Normal Breathing samples for each modeling approach. For machine learning models, we selected 110,000 Normal Breathing samples to maintain a balance between class distribution and model performance. For deep learning models, this was increased to 150,000, as deep architectures can handle slightly imbalanced distributions more effectively due to their enhanced learning capacity. Consequently, we retained more normal samples in deep learning models than in machine learning. Crucially, all apnea-related samples (OSA, CSA, MSA) were preserved in both approaches, ensuring no minority-class instances were removed and maintaining full representation of these critical events, as shown in Table~\ref{tab:Undersampling}.
\begin{table}[h!]
	\caption{Class Distribution Before and After Undersampling\label{tab:Undersampling}}
	\centering
	\renewcommand{\arraystretch}{1.5} 
	\begin{tabular}{|>{\centering\arraybackslash}p{0.67cm}|>{\centering\arraybackslash}p{1.575cm}|>{\centering\arraybackslash}p{2.81cm}|>{\centering\arraybackslash}p{2.365cm}|}
		\cline{3-4}
		\multicolumn{2}{c|}{} & \multicolumn{2}{c|}{\textbf{After Undersampling}}                       \\ \hline
		\textbf{Class} & \textbf{Original (\%)} & \textbf{Machine Learning (\%)} & \textbf{Deep Learning (\%)} \\ \hline
		Normal & 90 & 64 & 71 \\ \hline
		OSA & 5 & 17 & 14 \\ \hline
		CSA & 4 & 14 & 12 \\ \hline
		MSA & 1 & 4 & 3 \\ \hline
	\end{tabular}
\end{table}

\subsection{Feature Extraction and Selection}
During sleep apnea, heart rate decreases with reduced breathing and increases as breathing normalizes at the end of an apnea event. This behavior, known as Cyclical Variation of Heart Rate (CVHR), significantly affects R waves, ECG-Derived Respiration (EDR), and RR intervals, which reflect Heart Rate Variability (HRV) in ECG signals \cite{6}. Therefore, we first extract HRV from ECG window signals and then derive features from HRV. These features are categorized into three subsets: time-domain, frequency-domain, and non-linear features \cite{19}.
\begin{itemize}
	\item Time Domain Features:  Represent statistical measures of HRV. Extracted features include MeanNN, SDNN, RMSSD, SDSD, CVNN, CVSD, MedianNN, MadNN, MCVNN, IQRNN, Prc20NN, Prc80NN, pNN50, pNN20, MinNN, MaxNN, HTI, and TINN \cite{20}.
	\item Derived using spectral analysis of HRV, these describe power distribution across frequency bands. Extracted features include LF, HF, VHF, LFHF, LFn, HFn, and LnHF \cite{20}.
	
	\item Non-Linear Domain Features: Capture complex HRV patterns that cannot be identified through linear analysis. Extracted features include SD1, SD2, SD1\_SD2, S, CSI, CVI, CSI\_Modified, PIP, IALS, PSS, PAS, GI, SI, AI, PI, C1d, C1a, SD1d, SD1a, C2d, C2a, SD2d, SD2a, Cd, Ca, SDNNd, SDNNa, DFA\_alpha1, MFDFA\_alpha1\_Width, MFDFA\_alpha1\_Peak,HFD, KFD MFDFA\_alpha1\_Mean, MFDFA\_alpha1\_Max, MFDFA\_alpha1\_Delta, MFDFA\_alpha1\_Asymmetry, MFDFA\_alpha1\_Fluctuation, CMSEn, RCMSEn, CD, MFDFA\_alpha1\_Increment, ApEn, SampEn, ShanEn, FuzzyEn, MSEn,   and LZC \cite{20}.
\end{itemize}
In total, we extract 18 features for the WIN-11 approach and 72 features for WIN-61 and WIN-MIX. To mitigate the curse of dimensionality, we apply Recursive Feature Elimination (RFE), which iteratively removes less significant features, refining model performance by focusing on the most relevant data. Using RFE, we reduce the feature sets to 15 for WIN-11 and 40 for WIN-61 and WIN-MIX, thereby preventing overfitting, improving generalization, and reducing computational complexity.
\subsection{Machine Learning Algorithms} 
In this paper, we utilized six widely used machine learning algorithms to classify sleep apnea subtypes based on features extracted from HRV and selected using RFE. While our primary focus is energy-efficient deep learning, classical machine learning methods are inherently lightweight and can operate without graphics processing unit (GPU), making them well-suited for microsensor and wearable applications. However, these methods rely on feature extraction and generally require longer window sizes, which may limit real-time responsiveness. By incorporating these algorithms, we conduct a comprehensive analysis to determine the optimal balance between classification performance and energy consumption for sleep apnea subtype detection. The following subsections provide a brief overview of each algorithm.

\textbf{\textit{K-Nearest Neighbor (KNN)}} algorithm determines the class of a new data point by majority vote among the K closest training samples, making performance sensitive to the choice of K and outliers. It is straightforward but can be computationally expensive for large datasets.

\textbf{\textit{Support Vector Machine (SVM)}} algorithm identifies the hyperplane with the largest margin between classes, often requiring careful parameter selection (e.g., kernel function, regularization) for optimal performance. SVMs can handle nonlinear boundaries with kernel transformations.

\textbf{\textit{Logistic Regression (LR)}}  Uses a logistic function to model probabilities of each class, balancing interpretability and simplicity. By applying weights to input features, LR learns linear decision boundaries suitable for various classification problems.

\textbf{\textit{Decision Tree (DT)}} Recursively splits the feature space into distinct branches, ending in leaf nodes that represent class labels. While simple and interpretable, DTs can overfit unless properly regularized (e.g., via pruning)..

\textbf{\textit{Random Forest (RF)}} algorithm Builds multiple decision trees on different bootstrap samples of the data and aggregates their votes, reducing variance and overfitting. This ensemble approach is typically robust to noise and outliers.

\textbf{\textit{Gradient Boosting Decision Tree (GBDT)}} Sequentially trains decision trees, with each new tree correcting errors of the previous ones. This method can yield highly accurate ensembles at the cost of careful hyperparameter tuning and potentially longer training times.

\subsection{Hyperparameter Tuning}
Hyperparameters are configuration settings of machine learning algorithms that are defined before training. such as the number of neighbors in KNN, the strength of regularization in SVM, or the depth of trees in Random Forest or other tree-based algorithms. Therefore, hyperparameter tuning can have a significant impact on the model’s performance and their generalization ability and computational efficiency. For hyperparameter tuning, we have used the Bayesian optimization approach with 5-fold cross-validation; an efficient algorithm that searches through the space of possible hyperparameters by iteratively modeling and updating a probabilistic model of the objective function, thus providing near-optimal configurations for each of our machine learning models.

\subsection{Deep Learning Architectures}
In this paper, we evaluate eight deep learning architectures for sleep apnea subtype classification, each trained on 11-second ECG windows (WIN-11) to achieve 1-second resolution. Even the most lightweight deep networks (e.g., MobileNet-v1, SqueezeNet-1.1) still contain a considerable number of parameters and require significant energy, posing challenges for continuous operation on battery-powered devices. To address this, we developed a SleepLiteCNN optimized for 1D ECG signals, significantly reducing parameters while maintaining robust classification accuracy and lowering energy consumption. This approach aligns with our objective of real-time, low-power monitoring for wearable sleep apnea detection.

The following subsections provide a brief overview of each model, emphasizing their design, parameter efficiency, and suitability for energy-constrained environments.

\begin{figure*}[!hb]
	\centering
	\includegraphics[width=1\linewidth]{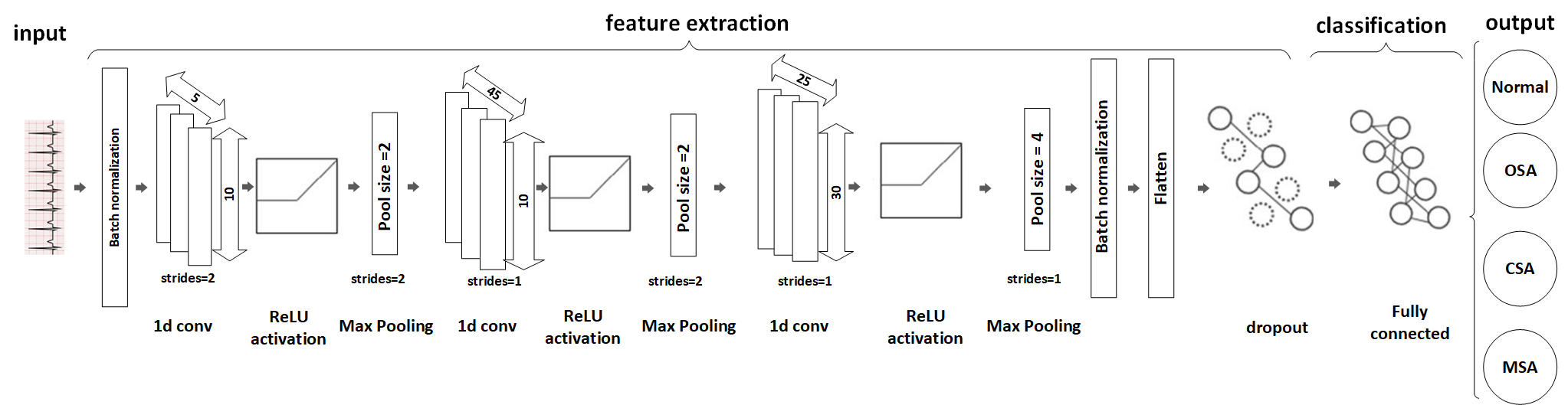}
		\caption{Architecture of the SleepLiteCNN Model for Sleep Apnea Subtype Classification.}
	\label{fig_3}
\end{figure*}

\textbf{\textit{VGG-11}} is the simplest variant of the VGG family, chosen for its well‐established performance and straightforward, uniform structure. Despite having more parameters than some modern architectures, it serves as a strong baseline and is relatively easy to adapt for 1D ECG signals.

\textbf{\textit{MobileNet-v1}} is designed for resource‐constrained platforms, making it highly suitable for wearable ECG classification. By using depthwise separable convolutions, it drastically reduces parameter count (400K in our implementation) without severely compromising accuracy.

\textbf{\textit{AlexNet}} famous for its success in the 2012 ImageNet competition, consists of five convolutional layers followed by three fully connected layers, all using ReLU activations. It introduced important innovations like dropout and local response normalization, with our version containing 64 million parameters, making it an effective model for tasks like sleep apnea classification from ECG data.

\textbf{\textit{SqueezeNet-1.1}}specifically aims for AlexNet‐level accuracy with fewer parameters, making it ideal for deployment on devices with limited memory or power. Its "fire modules" (1x1 and 3x1 convolutions) further minimize computational overhead for 1D data.

\textbf{\textit{Inception-v3}} employs parallel convolution paths (“Inception modules”) to capture diverse features at multiple scales. Though more complex than MobileNet or SqueezeNet, it offers a useful benchmark for how multi‐scale feature extraction benefits ECG classification.

\textbf{\textit{EfficientNet-B2}} leverages mobile inverted bottleneck blocks and squeeze‐excitation for high accuracy at relatively low parameter counts (159K in our setup). Its systematic scaling strategy provides insight into how effectively it balances performance and efficiency on 1D signals.

\textbf{\textit{ResNet-18}} includes residual connections that help mitigate vanishing gradients, which is especially helpful in deeper networks. As the lightest ResNet variant, it combines moderate depth (18 layers) with a manageable parameter count (3.9M) for our ECG classification task.

\textbf{\textit{SleepLiteCNN}} was developed because even the most lightweight standard architectures, such as MobileNet‐v1 and SqueezeNet‐1.1, still have a significant number of parameters and, consequently, high energy requirements. This presents a major challenge for battery-powered wearables that need to operate continuously. Our goal was to design a network that maintains robust classification performance while drastically reducing energy consumption for real-time ECG analysis. After extensive experimentation and iterative refinement of hyperparameters, layer configurations, and filter counts, we developed a network (see Fig.~\ref{fig_3}) that comprises three convolutional layers with 5, 45, and 25 filters respectively—each followed by ReLU activations and max-pooling. The network also incorporates input batch-normalization, a dropout layer, and a final softmax layer with four neurons. With roughly 39K parameters, this SleepLiteCNN achieves competitive accuracy in apnea subtype detection while significantly lowering computational and energy demands, making it ideal for energy-constrained wearable devices.

\section{Results}
This section presents the experimental results, including the performance evaluation and comparison of various machine learning and deep learning models for classifying sleep apnea subtypes using ECG signals. Additionally, we analyze the energy consumption of deep learning models and examine the impact of quantization techniques. To assess hardware efficiency, we synthesize the most accurate and energy-efficient deep learning model on an FPGA using High-Level Synthesis (HLS). Given the class imbalance in our dataset, we use F1-macro as the primary evaluation metric to ensure balanced performance across minority and majority classes. We also report macro-averaged precision and recall to further validate model effectiveness across all apnea subtypes.

Table~\ref{tab:ml_performance} summarizes the performance of machine learning models across different windowing approaches (WIN-11, WIN-61, WIN-MIX). The RF model demonstrated the best performance with WIN-61, achieving 95\% accuracy and a 94\% F1-score. Similarly, GBDT and KNN performed well, both exceeding 90\% accuracy and F1-score with WIN-61 and WIN-MIX. Due to their ability to capture complex decision boundaries and mitigate overfitting using multiple weak learners, RF and GBDT are particularly well-suited for ensemble approaches. In contrast, the WIN-11 approach yielded significantly lower performance across all models, highlighting that longer time windows provide richer features for machine learning-based classification. While machine learning models are generally more energy-efficient than deep learning models, their reliance on longer time windows (WIN-61) for optimal performance limits their suitability for real-time apnea detection in wearable applications.

\begin{table*}[!ht]
	\captionsetup{width=0.5\textwidth}
\begin{minipage}[b]{0.55\textwidth}
	\caption{Performance Metrics of Machine Learning Algorithms Using Different Time Window Approaches for Sleep Apnea Classification\label{tab:ml_performance}}
	\centering
	\renewcommand{\arraystretch}{1.5} 
	\begin{tabular}{
			>{\centering\arraybackslash}p{0.5cm}|
			>{\centering\arraybackslash}p{1cm}|
			>{\centering\arraybackslash}p{1.1cm}|
			>{\centering\arraybackslash}p{1cm}|
			>{\centering\arraybackslash}p{1.1cm}|
			>{\centering\arraybackslash}p{1cm}|
			>{\centering\arraybackslash}p{1.1cm}|
		}
		\cline{2-7}
		&\multicolumn{2}{c|}{\textbf{WIN-11}}                       & \multicolumn{2}{c|}{\textbf{WIN-MIX}}                      & \multicolumn{2}{c|}{\textbf{WIN-61}}                       \\ \hline
		\multicolumn{1}{|c|}{\textbf{Model}} & \multicolumn{1}{>{\centering\arraybackslash}p{0.95cm}|}{\textbf{Accuracy (\%)}} & \textbf{F1-Score (\%)} & \multicolumn{1}{>{\centering\arraybackslash}p{0.95cm}|}{\textbf{Accuracy (\%)}} & \textbf{F1-Score (\%)} & \multicolumn{1}{>{\centering\arraybackslash}p{0.95cm}|}{\textbf{Accuracy (\%)}} & \textbf{F1-Score (\%)} \\ \hline
				\multicolumn{1}{|c|}{RF}            & \multicolumn{1}{c|}{69}                & 48                & \multicolumn{1}{c|}{94}                & 93                & \multicolumn{1}{c|}{95}                & 94                \\ \hline
		
		\multicolumn{1}{|c|}{GBDT}          & \multicolumn{1}{c|}{68}                & 43                & \multicolumn{1}{c|}{93}                & 91                & \multicolumn{1}{c|}{95}                & 94                \\ \hline

		\multicolumn{1}{|c|}{KNN}           & \multicolumn{1}{c|}{63}                & 45                & \multicolumn{1}{c|}{94}                & 93                & \multicolumn{1}{c|}{95}                & 94                \\ \hline
	
		\multicolumn{1}{|c|}{DT}            & \multicolumn{1}{c|}{57}                & 40                & \multicolumn{1}{c|}{81}                & 74                & \multicolumn{1}{c|}{84}                & 80                \\ \hline
	
		\multicolumn{1}{|c|}{SVM}           & \multicolumn{1}{c|}{47}                & 38               & \multicolumn{1}{c|}{91}                & 87                & \multicolumn{1}{c|}{93}                & 91                \\ \hline
	
		\multicolumn{1}{|c|}{LR}            & \multicolumn{1}{c|}{65}                & 22                & \multicolumn{1}{c|}{66}                & 33                & \multicolumn{1}{c|}{66}                & 34                \\ \hline
	\end{tabular}
\end{minipage}
	\hfill
	\begin{minipage}[b]{0.45\textwidth}
	\caption{Deep Learning Classification Results for Sleep Apnea Subtypes\label{tab:cnn_performance}}
		\centering
		\renewcommand{\arraystretch}{1.5} 
		\begin{tabular}{|>{\centering\arraybackslash}p{1.8cm}|>{\centering\arraybackslash}p{0.95cm}|>{\centering\arraybackslash}p{1.1cm}|>{\centering\arraybackslash}p{0.95cm}|>{\centering\arraybackslash}p{0.8cm}|}
			\hline
			\textbf{Model} & \textbf{Accuracy (\%)} & \textbf{F1-Score (\%)} & \textbf{Precision (\%)} & \textbf{Recall (\%)} \\ \hline
			SleepLiteCNN & 95 & 92 & 91 & 94 \\ \hline
			VGG-11 & 96 & 95 & 95 & 94 \\ \hline
			MobileNet-v1 & 96 & 94 & 94 & 94 \\ \hline
			AlexNet & 95 & 93 & 94 & 93 \\ \hline
			SqueezeNet-1.1 & 94 & 92 & 91 & 93 \\ \hline
			Inception-v3 & 93 & 88 & 90 & 87 \\ \hline
			EfficientNet-B2 & 84 & 74 & 73 & 75 \\ \hline
			ResNet-18 & 73 & 35 & 62 & 34 \\ \hline
		\end{tabular}
	\end{minipage}
\end{table*}

\begin{figure*}[!ht]
	\centering
	\subfloat[]{\includegraphics[width=0.45\linewidth]{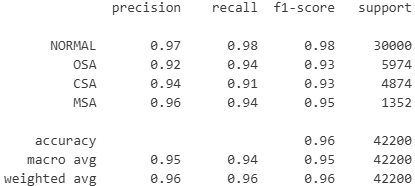}%
		\label{fig_first_case}}
	\hfil
	\subfloat[]{\includegraphics[width=0.32\linewidth]{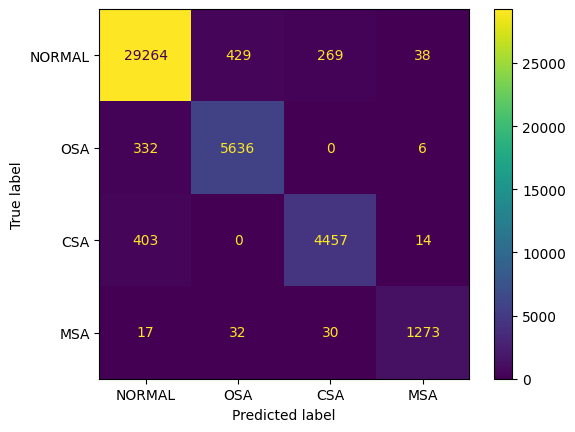}%
		\label{fig_second_case}}
	
	\caption{Classification report (a) and confusion matrix (b) for VGG-11, our best-performing model for sleep apnea subtype classification.}
	\label{vgg11}
\end{figure*}

\begin{table*}[t]
	\captionsetup{width=1\textwidth}
	\begin{minipage}[b]{0.70\textwidth}
		\caption{Energy Consumption and Performance Comparison of Deep Learning Models Before and After 8-bit Quantization.\label{tab:model_comparison}}
		\centering
		\renewcommand{\arraystretch}{1.5} 
		\begin{tabular}{%
				>{\centering\arraybackslash}p{0.5cm}
				>{\centering\arraybackslash}p{0.75cm}|
				>{\centering\arraybackslash}p{1cm}|
				>{\centering\arraybackslash}p{1.1cm}|
				>{\centering\arraybackslash}p{1cm}|
				>{\centering\arraybackslash}p{1cm}|
				>{\centering\arraybackslash}p{1.1cm}|
				>{\centering\arraybackslash}p{1cm}|
			}
			\cline{3-8}
			& & \multicolumn{3}{c|}{\textbf{Full Precision}} & \multicolumn{3}{c|}{\textbf{8-bit Quantization}} \\
			\hline
			\multicolumn{1}{|c|}{\textbf{Model}} & \textbf{Total Params} & \textbf{Accuracy (\%)} & \textbf{F1-Score (\%)} & \textbf{Energy ($\mu J$)} & \textbf{Accuracy (\%)} & \textbf{F1-Score (\%)} & \textbf{Energy ($\mu J$)} \\
			\hline
			\multicolumn{1}{|c|}{SleepLiteCNN}       & 39K   & 95  & 92  & 25.76    & 94  & 91  & 1.80    \\ \hline
			\multicolumn{1}{|c|}{VGG-11}         & 112M  & 96  & 95  & 3077.06  & 96  & 94  & 2730.88 \\ \hline
			\multicolumn{1}{|c|}{MobileNet-v1}   & 400K  & 96  & 94  & 182.34   & 95  & 92  & 11.46   \\ \hline
			\multicolumn{1}{|c|}{AlexNet}        & 62M   & 95  & 93  & 987.53   & 95  & 93  & 824.26  \\ \hline
			\multicolumn{1}{|c|}{SqueezeNet-1.1} & 354K  & 94  & 92  & 174.62   & 94  & 92  & 174.11  \\ \hline
			\multicolumn{1}{|c|}{Inception-v3}   & 1.5M  & 93  & 88  & 802.42   & 92  & 89  & 800.16  \\ \hline
			\multicolumn{1}{|c|}{EfficientNet-B2}& 159K  & 84  & 74  & 5.11     & 83  & 74  & 4.89    \\ \hline
			\multicolumn{1}{|c|}{ResNet-18}      & 3.9M  & 73  & 35  & 1111.63  & 71  & 26  & 71.25   \\ \hline
		\end{tabular}
	\end{minipage}
	\hfill
	\begin{minipage}[b]{0.30\textwidth}
	\caption{FPGA Resource Utilization for Our SleepLiteCNN Model Before and After 8-bit Quantization\label{tab:Utilization}}
	\centering
	\renewcommand{\arraystretch}{1.5}
	\begin{tabular}{c|cc|}
		\cline{2-3}
		& \multicolumn{2}{c|}{\textbf{Utilization (\%)}}    \\ \hline
		\multicolumn{1}{|c|}{\textbf{Resource} } & \multicolumn{1}{c|}{\textbf{Full Precision} } &  \textbf{8-bit Quantization} \\ \hline
		\multicolumn{1}{|c|}{LUT} & \multicolumn{1}{c|}{31.56} & 23.21 \\ \hline
		\multicolumn{1}{|c|}{FF} & \multicolumn{1}{c|}{33.2} & 23.92 \\ \hline
		\multicolumn{1}{|c|}{BRAM} & \multicolumn{1}{c|}{22.47} & 19.32 \\ \hline
		\multicolumn{1}{|c|}{DSP} & \multicolumn{1}{c|}{26.08} & 13.38 \\ \hline
	\end{tabular}
\end{minipage}

\end{table*}

\begin{figure*}[!ht]
	\centering
	\subfloat[]{\includegraphics[width=0.5\linewidth]{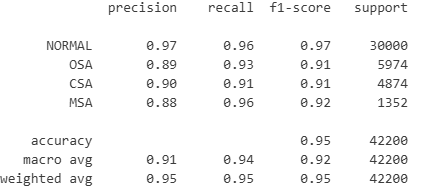}%
		\label{cnn1}}
	\hfil
	\subfloat[]{\includegraphics[width=0.32\linewidth]{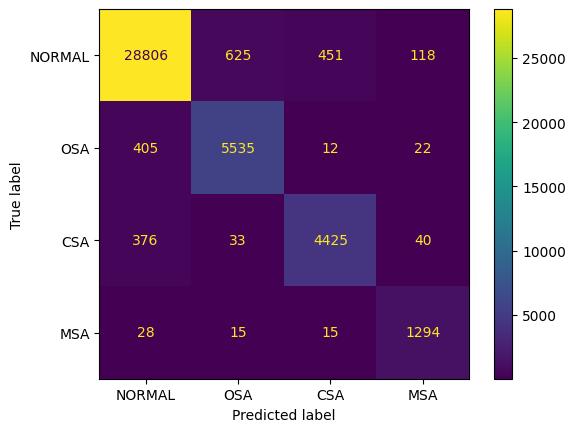}%
		\label{cnn2}}
\vfil
	\subfloat[]{\includegraphics[width=1\linewidth]{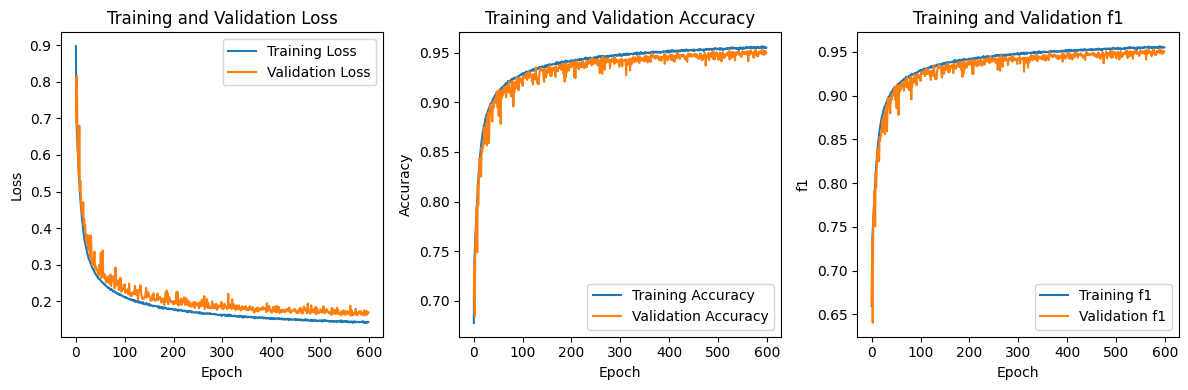}%
	\label{cnn_v}}
	\caption{Performance evaluation of the proposed SleepLiteCNN model. (a) Classification report summarizing precision, recall, and F1-score for each sleep apnea subtype. (b) Confusion matrix illustrating per-class prediction accuracy. (c) Training and validation curves for loss, accuracy, and F1-score over 600 epochs, demonstrating stable convergence and generalization.}
	
	\label{cnn}
\end{figure*}

\begin{figure*}[!ht]

	\centering
	\subfloat[]{\includegraphics[width=0.5\linewidth]{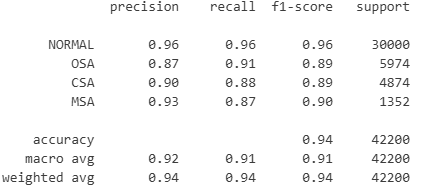}%
		\label{cnnq1}}
	\hfil
	\subfloat[]{\includegraphics[width=0.32\linewidth]{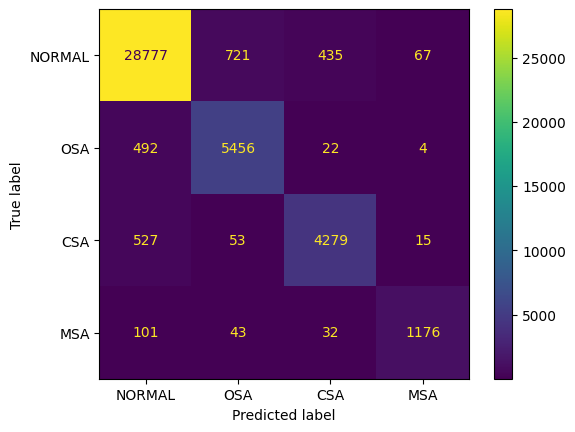}%
		\label{cnnq2}}
		
		\vfil
		\subfloat[]{\includegraphics[width=1\linewidth]{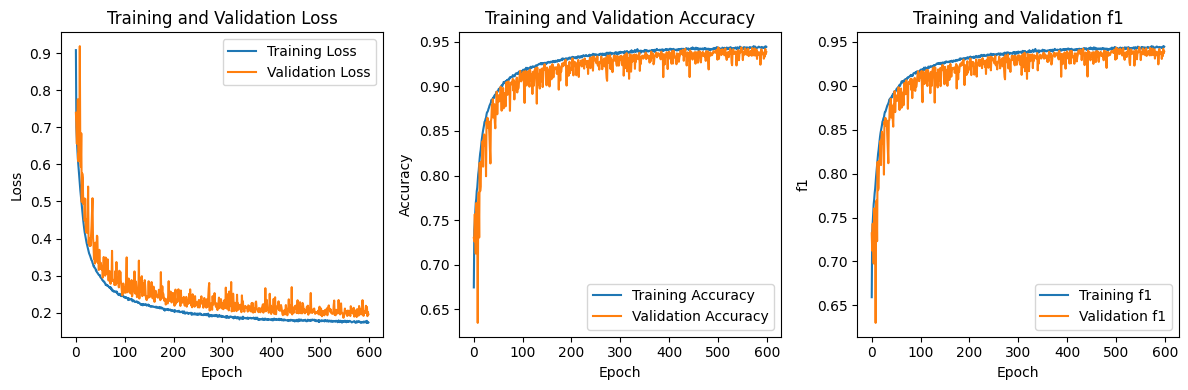}%
			\label{cnn_vq}}
	\caption{Performance evaluation of the proposed SleepLiteCNN model after 8-bit quantization. (a) Classification report summarizing precision, recall, and F1-score for each sleep apnea subtype. (b) Confusion matrix illustrating per-class prediction accuracy. (c) Training and validation curves for loss, accuracy, and F1-score over 600 epochs, demonstrating stable convergence and generalization. }
	
	\label{cnnq}
\end{figure*}

Table~\ref{tab:cnn_performance} summarizes the classification performance of deep learning models trained with the WIN-11 approach, enabling 1-second resolution. Most models achieved high accuracy, with VGG-11 and MobileNet-v1 delivering the best results at 96\% accuracy and 94-95\% F1-score. Our SleepLiteCNN also performed well, attaining 95\% accuracy and 92\% F1-score, while consuming significantly less energy than larger architectures. To provide a more detailed per-class analysis, Fig.~\ref{vgg11} presents the classification report and confusion matrix for VGG-11, the top-performing model. These results confirm VGG-11’s strong performance across all sleep apnea subtypes, demonstrating its ability to accurately classify each category.

In this paper, we applied quantization techniques to optimize energy consumption in deep learning models. Quantization reduces the precision of neural network weights and activations, typically from 32-bit floating-point to lower bit-width representations \cite{21}, significantly reducing model size and computational requirements while maintaining accuracy. To implement this, we used QKeras, an extension of Keras that supports quantization-aware training and provides tools for estimating energy consumption, which we leveraged for evaluation under the 45nm process \cite{21}. We specifically applied 8-bit quantization, as this format is widely supported by most hardware platforms, ensuring compatibility across embedded and wearable devices. Table~\ref{tab:model_comparison} compares energy consumption and classification performance before and after quantization. Our SleepLiteCNN demonstrated the lowest energy consumption, requiring only 1.80~$\mu$J post-quantization while maintaining 94\% accuracy. In contrast, larger models like VGG-11 consumed significantly more energy (2730.88~$\mu$J) despite achieving slightly higher accuracy. This underscores the trade-off between classification performance and energy efficiency. Additionally, MobileNet-v1 and SqueezeNet-1.1 delivered strong performance while consuming substantially less energy than larger architectures, making them well-suited for wearable applications.

Considering the balance between accuracy and energy efficiency, we selected our SleepLiteCNN model for FPGA synthesis to demonstrate its feasibility for resource-constrained devices while maintaining high classification accuracy. We utilized hls4ml, a tool that enables  HLS \cite{22}, to convert our model into an FPGA implementation. FPGAs offer low-power, high-performance execution, making them ideal for wearable applications. By deploying our SleepLiteCNN on an Artix-7 FPGA (xc7a200tfbg484-3), we evaluated its hardware resource utilization. As shown in Table~\ref{tab:Utilization}, quantization significantly reduced FPGA resource consumption, leading to a 26.5\% decrease in LUT utilization (from 31.56\% to 23.21\%), a 27.9\% reduction in FF usage (from 33.2\% to 23.92\%), a 14.0\% decrease in BRAM utilization (from 22.47\% to 19.32\%), and a 48.7\% reduction in DSP usage (from 26.08\% to 13.38\%). These results confirm that quantization not only enhances deployment efficiency but also validates the SleepLiteCNN as a viable solution for real-time, low-power wearable devices.

\begin{table*}[t]
	\centering
	\caption{Comparative Analysis of State-of-the-Art Sleep Apnea Detection Methods}
	\label{tab:sleep_stage_models}
	\renewcommand{\arraystretch}{1.5}
	\begin{tabular}{|>{\centering\arraybackslash}p{3.1cm}|>{\centering\arraybackslash}p{0.65cm}|>{\centering\arraybackslash}p{3.5cm}|>{\centering\arraybackslash}p{3.5cm}|>{\centering\arraybackslash}p{0.85cm}|>{\centering\arraybackslash}p{1.7cm}|>{\centering\arraybackslash}p{1.7cm}|}
		\hline
		\textbf{Model} & \textbf{Signal} & \textbf{Temporal Resolution} & \textbf{Class Labels} & \textbf{Dataset} & \textbf{Accuracy (\%)} & \textbf{F1-Score (\%)} \\
		\hline
		RF \cite{2} & EEG & - & Normal, OSA, CSA & Tianjin & 88.99 & 87 \\ \hline
		Modified LeNet5-K5 \cite{4} & PSG & 1 Second (60-second window) & Normal, OSA, CSA, MSA, Hypopnea  & OSAUD & 94.26 & 94.6 \\ \hline
		1D-CNN \cite{9} & ECG & 1 Second (11-second window) & Normal, Apnea & UCDDB & 99.56 & 95 \\ \hline
		ZFNet-BiLSTM \cite{12} & ECG & 1 Minute & Normal, Apnea & AED & 88.13 & 84.04 \\ \hline
		CNNLSTM \cite{13} & ECG & 1 Minute & Normal, Apnea & AED & 86.25 & 87.68 \\ \hline
		subspace KNN \cite{15} & EEG & 10 to 20 Second & Normal, OSA, CSA, MSA, O-HYP, C-HYP, M-HYP & UCDDB & 81.39 & 51 \\ \hline
		Multiscale 1D-CNN \cite{23} & HRV & 1 Minute &  Normal, Apnea & AED & 96 & - \\ \hline
		
		Our best model (VGG-11) & ECG & 1 Second (11-second window) & Normal, OSA, CSA, MSA & UCDDB & 96 & 95 \\ 
		\hline
			Our SleepLiteCNN  & ECG & 1 Second (11-second window) & Normal, OSA, CSA, MSA & UCDDB & 95 & 92 \\ 
		\hline
	\end{tabular}
\end{table*}
Fig.\ref{cnn} and Fig.\ref{cnnq} illustrate the classification performance of the proposed SleepLiteCNN model before and after 8-bit quantization, respectively. Subfigures (a) in both figures present the per-class precision, recall, and F1-score metrics, showing strong and balanced performance across all four sleep apnea subtypes. Subfigures (b) display the corresponding confusion matrices, confirming the model’s ability to distinguish each class with high fidelity—particularly in distinguishing CSA and MSA from OSA and normal breathing.
Subfigures (c) in both figures show the training and validation curves for loss, accuracy, and macro F1-score over 600 epochs. The convergence trends remain smooth and stable in both settings, indicating strong generalization and no signs of overfitting.
After quantization (Fig.~\ref{cnnq}), SleepLiteCNN shows only a marginal drop in overall accuracy (from 95\% to 94\%) and macro F1-score (from 0.92 to 0.91), while maintaining consistent learning dynamics. This minimal degradation confirms the robustness of the architecture under reduced precision and highlights its practical suitability for energy-constrained, real-time deployment on wearable devices.

Table~\ref{tab:sleep_stage_models} compares our best models with state-of-the-art sleep apnea detection approaches. VGG-11 achieved 96\% accuracy and 95\% F1-score, offering full apnea subtype classification (OSA, CSA, MSA, and normal breathing) with 1-second resolution using an 11-second window (WIN-11). Unlike prior studies that primarily focus on binary classification, our method provides detailed diagnostic insights, facilitating more precise treatment decisions. While some studies employed PSG or EEG, our ECG-based approach is better suited for wearable devices. Although the 1D-CNN model \cite{9} attained a higher accuracy (99.56\%), it only differentiates between apnea and normal breathing, limiting its clinical applicability.

Several prior studies, such as \cite{24} and \cite{25}, have investigated FPGA implementations for sleep apnea classification, but their methodologies differ significantly from ours, making direct comparisons challenging. Many relied on EEG and SpO\textsubscript{2} rather than ECG, focused solely on binary classification, and lacked energy consumption metrics or optimization techniques like quantization. Additionally, their models used longer temporal resolutions (30s and 10s), restricting comparability with our 1-second approach. The study in \cite{9} does provide energy consumption estimates, but their evaluation was conducted on 28nm technology, whereas ours is based on 45nm. They also employed pruning instead of quantization while still focusing on binary classification. Among their models, the closest to ours is a 1D-CNN with 5\% pruning, which consumes 2.18~$\mu$J per inference. In contrast, our optimized quantized model achieves significantly lower energy consumption while maintaining high classification performance across all apnea subtypes, making it better suited for real-time, energy-efficient wearable deployment.

\section{Limitations and Future Directions}

Despite these promising findings, this study is limited by the scarcity and expense of PSG data, which can be difficult to obtain. Additionally, most publicly available datasets are labeled in 30-second or 1-minute intervals rather than at finer resolutions. These datasets often place greater emphasis on OSA compared to CSA and MSA, leading to class imbalance that can impede accurate subtype detection.

Future work will aim to address these challenges by evaluating the proposed models with real-world wearable ECG data, thereby ensuring robust performance in less controlled conditions. Second, as more data becomes available, exploring subject-specific or leave-one-subject-out analyses could further reduce overfitting and offer insights into individual-level variability. In addition, transfer learning may enhance model adaptability, especially for underrepresented apnea subtypes. Finally, investigating alternative windowing strategies could improve temporal resolution for earlier detection, moving the field closer to real-time monitoring in wearable applications.

\section{Conclusion}

This paper explored the challenge of achieving accurate, real-time sleep apnea subtype detection within the tight energy and hardware constraints of wearable devices. By systematically evaluating a diverse set of machine learning and deep learning models on 1-second single-lead ECG data, we demonstrated that traditional high-accuracy models are often impractical for edge deployment due to their size and energy demands.
To address this, we introduced SleepLiteCNN, a custom-designed CNN that bridges the gap between performance and efficiency. The model consistently delivered strong classification results across all apnea subtypes while meeting strict energy and resource limitations, confirmed through 8-bit quantization and FPGA synthesis.
These results not only validate the viability of real-time ECG-based apnea monitoring on wearables, but also highlight the importance of designing models specifically for the constraints of embedded hardware. SleepLiteCNN serves as a reproducible and hardware-conscious baseline for future work, enabling broader adoption of real-time, at-home sleep disorder diagnostics.
\section*{Code Availability}
The source code and scripts to reproduce the experiments presented in this paper, including training, quantization, and FPGA synthesis of SleepLiteCNN, are available at: \href{https://github.com/zahraaayii/SleepLiteCNN-Energy-Efficient-Sleep-Apnea-Subtype-Classification}{github.com/zahraaayii/SleepLiteCNN-Energy-Efficient-Sleep-Apnea-Subtype-Classification}.

\vspace{11pt}

\vfill


\begin{thebibliography}{1}
\bibliographystyle{IEEEtran}
\bibitem{1}
K. Pavlova and V. Latreille, 
"Sleep Disorders," 
\textit{The American Journal of Medicine}, vol. 132, no. 3, pp. 292–299, Mar. 2019, doi: 10.1016/j.amjmed.2018.09.021.

\bibitem{2}
X. Zhao, X. Wang, T. Yang, S. Ji, H. Wang, J. Wang, Y. Wang, and Q. Wu, 
"Classification of sleep apnea based on EEG sub-band signal characteristics," 
\textit{Scientific Reports}, vol. 11, no. 1, Mar. 2021, doi: 10.1038/s41598-021-85138-0.

\bibitem{3}
A. Sabil, C. Marien, M. LeVaillant, G. Baffet, N. Meslier, and F. Gagnadoux, 
"Diagnosis of sleep apnea without sensors on the patient's face," 
\textit{Journal of Clinical Sleep Medicine}, vol. 16, no. 7, pp. 1161–1169, Jul. 2020, doi: 10.5664/jcsm.8460.

\bibitem{4}
A. De and E. Priya, 
"Sleep Apnea sub-type detection from Polysomnography signals," 
in \textit{2024 IEEE International Conference on Interdisciplinary Approaches in Technology and Management for Social Innovation (IATMSI)}, Mar. 2024, pp. 1–6, doi: 10.1109/iatmsi60426.2024.10503128.

\bibitem{5}
J. V. Rundo and R. Downey, 
"Polysomnography," 
in \textit{Clinical Neurophysiology: Basis and Technical Aspects}, Elsevier, 2019, pp. 381–392, doi: 10.1016/b978-0-444-64032-1.00025-4.

\bibitem{6}
T. Penzel, G. B. Moody, R. G. Mark, A. L. Goldberger, and J. H. Peter, 
"The apnea-ECG database," 
in \textit{Computers in Cardiology 2000. Vol.27 (Cat. 00CH37163)}, IEEE, pp. 255–258, doi: 10.1109/cic.2000.898505.

\bibitem{7}
A. John, K. K. Nundy, B. Cardiff, and D. John, 
"Multimodal Multiresolution Data Fusion Using Convolutional Neural Networks for IoT Wearable Sensing," 
\textit{IEEE Transactions on Biomedical Circuits and Systems}, vol. 15, no. 6, pp. 1161–1173, Dec. 2021, doi: 10.1109/tbcas.2021.3134043.

\bibitem{8}
A. John, K. K. Nundy, B. Cardiff, and D. John, 
"SomnNET: An SpO2 Based Deep Learning Network for Sleep Apnea Detection in Smartwatches," 
in \textit{2021 43rd Annual International Conference of the IEEE Engineering in Medicine \& Biology Society (EMBC)}, Nov. 2021, pp. 1961–1964, doi: 10.1109/embc46164.2021.9631037.

\bibitem{9}
A. John, B. Cardiff, and D. John, 
"A 1D-CNN Based Deep Learning Technique for Sleep Apnea Detection in IoT Sensors," 
in \textit{2021 IEEE International Symposium on Circuits and Systems (ISCAS)}, May 2021, doi: 10.1109/iscas51556.2021.9401300.

\bibitem{10}
E. Urtnasan, J.-U. Park, E.-Y. Joo, and K.-J. Lee, 
"Automated Detection of Obstructive Sleep Apnea Events from a Single-Lead Electrocardiogram Using a Convolutional Neural Network," 
\textit{Journal of Medical Systems}, vol. 42, no. 6, Apr. 2018, doi: 10.1007/s10916-018-0963-0.

\bibitem{11}
D. Dey, S. Chaudhuri, and S. Munshi, 
"Obstructive sleep apnoea detection using convolutional neural network based deep learning framework," 
\textit{Biomedical Engineering Letters}, vol. 8, no. 1, pp. 95–100, Dec. 2017, doi: 10.1007/s13534-017-0055-y.

\bibitem{12}
M. Bahrami and M. Forouzanfar, 
"Sleep Apnea Detection From Single-Lead ECG: A Comprehensive Analysis of Machine Learning and Deep Learning Algorithms," 
\textit{IEEE Transactions on Instrumentation and Measurement}, vol. 71, pp. 1–11, 2022, doi: 10.1109/tim.2022.3151947.

\bibitem{13}
A. Sheta, H. Turabieh, T. Thaher, J. Too, M. Mafarja, M. S. Hossain, and S. R. Surani, 
"Diagnosis of Obstructive Sleep Apnea from ECG Signals Using Machine Learning and Deep Learning Classifiers," 
\textit{Applied Sciences}, vol. 11, no. 14, p. 6622, Jul. 2021, doi: 10.3390/app11146622.

\bibitem{14}
L. Cen, Z. L. Yu, T. Kluge, and W. Ser, 
"Automatic System for Obstructive Sleep Apnea Events Detection Using Convolutional Neural Network," 
in \textit{2018 40th Annual International Conference of the IEEE Engineering in Medicine and Biology Society (EMBC)}, Jul. 2018, pp. 3975–3978, doi: 10.1109/embc.2018.8513363.

\bibitem{15}
A. Chatterjee and N. D. Jana, 
"Classification of Sleep Apnea Event Type Using Imbalanced Labelled EEG Signal," 
in \textit{2022 IEEE Region 10 Symposium (TENSYMP)}, Jul. 2022, pp. 1–6, doi: 10.1109/tensymp54529.2022.9864566.

\bibitem{16}
M. Cheng, W. J. Sori, F. Jiang, A. Khan, and S. Liu, 
"Recurrent Neural Network Based Classification of ECG Signal Features for Obstruction of Sleep Apnea Detection," 
in \textit{2017 IEEE International Conference on Computational Science and Engineering (CSE) and IEEE International Conference on Embedded and Ubiquitous Computing (EUC)}, Jul. 2017, pp. 199–202, doi: 10.1109/cse-euc.2017.220.

\bibitem{24}
O. Hassan et al., 
"Energy Efficient Deep Learning Inference Embedded on FPGA for Sleep Apnea Detection," 
\textit{Journal of Signal Processing Systems}, vol. 94, no. 6, pp. 609–619, Jan. 10, 2022, 
doi: 10.1007/s11265-021-01722-7.

\bibitem{25}
Md. S. Alam, Y. Siddiqui, M. Hasan, O. Farooq, and Y. Himeur, 
"Energy-Efficient FPGA Based Sleep Apnea Detection Using EEG Signals," 
\textit{IEEE Access}, vol. 12, pp. 40182–40195, 2024, 
doi: 10.1109/ACCESS.2024.3374223.

\bibitem{17}
W. McNicholas, L. Doherty, S. Ryan, J. Garvey, P. Boyle, and E. Chua, 
"St. Vincent's University Hospital / University College Dublin Sleep Apnea Database," 
\textit{physionet.org}, 2004. [Online]. Available: https://physionet.org/content/ucddb/. doi: 10.13026/C26C7D.




\bibitem{19}
T. Pham, Z. J. Lau, S. H. A. Chen, and D. Makowski, 
"Heart Rate Variability in Psychology: A Review of HRV Indices and an Analysis Tutorial," 
\textit{Sensors}, vol. 21, no. 12, p. 3998, Jun. 2021, doi: 10.3390/s21123998.

\bibitem{20}
D. Makowski, T. Pham, Z. J. Lau, J. C. Brammer, F. Lespinasse, H. Pham, C. Schölzel, and S. H. A. Chen, 
"NeuroKit2: A Python toolbox for neurophysiological signal processing," 
\textit{Behavior Research Methods}, vol. 53, no. 4, pp. 1689–1696, Feb. 2021, doi: 10.3758/s13428-020-01516-y.

\bibitem{21}
C. N. Coelho, A. Kuusela, S. Li, H. Zhuang, J. Ngadiuba, T. K. Aarrestad, V. Loncar, M. Pierini, A. A. Pol, and S. Summers, 
"Automatic heterogeneous quantization of deep neural networks for low-latency inference on the edge for particle detectors," 
\textit{Nature Machine Intelligence}, vol. 3, no. 8, pp. 675–686, Jun. 2021, doi: 10.1038/s42256-021-00356-5.

\bibitem{22}
J. Duarte, S. Han, P. Harris, S. Jindariani, E. Kreinar, B. Kreis, J. Ngadiuba, M. Pierini, R. Rivera, N. Tran, and Z. Wu, 
"Fast inference of deep neural networks in FPGAs for particle physics," 
\textit{Journal of Instrumentation}, vol. 13, no. 07, pp. P07027–P07027, Jul. 2018, doi: 10.1088/1748-0221/13/07/p07027.

\bibitem{23}
Q. Shen, H. Qin, K. Wei, and G. Liu, 
"Multiscale Deep Neural Network for Obstructive Sleep Apnea Detection Using RR Interval From Single-Lead ECG Signal," 
\textit{IEEE Transactions on Instrumentation and Measurement}, vol. 70, pp. 1–13, 2021, 
doi: 10.1109/TIM.2021.3062414.



\end{thebibliography}
\end{document}